# Flow-driven magnetic microcatheter for superselective arterial embolization

**Authors:**

Lucio Pancaldi[1], Ece Özelçi [1], Mehdi Ali Gadiri[1], Julian Raub[1], Pascal John Mosimann[2]*, and Mahmut Selman Sakar[1]*

**Affiliations:**

[1] Institute of Mechanical Engineering, Ecole Polytechnique Fédérale de Lausanne (EPFL), 1015 Lausanne, Switzerland

[2] Neuroradiology Division, Toronto Western Hospital, Toronto, Ontario, Canada

*Corresponding authors:

Pascal John Mosimann and Mahmut Selman Sakar

Email: pascal.mosimann@gmail.com and selman.sakar@epfl.ch

**Abstract**

Minimally invasive interventions performed inside brain vessels with the synergistic use of microcatheters pushed over guidewires have revolutionized the way aneurysms, stroke, arteriovenous malformations, brain tumors and other cerebrovascular conditions are being treated. However, a significant portion of the brain vasculature remains inaccessible from within because the conventional catheterization technique based on transmitting forces from the proximal to the distal end of the instruments imposes stringent constraints on their diameter and stiffness. Here we overcome this mechanical barrier by microengineering a new class of ultraminiaturized magnetic microcatheters in the form of an inflatable flat tube, making them extremely flexible and capable of harnessing the kinetic energy of blood flow for endovascular navigation. We introduce a compact and versatile magnetic steering platform that is compatible with conventional bi-plane fluoroscope imaging, and demonstrate for the first time safe and effortless navigation and tracking of hard-to-reach, distal, tortuous arteries that are as small as 180 µm in diameter with a curvature radius as small as 0.69 mm. Furthermore, we demonstrate the superselective infusion of contrast and embolic liquid agents, all in a porcine model. These results pave the way to reach, diagnose, and treat currently inaccessible distal arteries that may be at risk of bleeding or feeding a tumor. Our endovascular technology can also be used to selectively target tissues for drug or gene delivery from within the arteries, not only in the central and peripheral nervous system but almost any other organ system, with improved accuracy, speed and safety.

# INTRODUCTION

The conventional catheterization procedure involves the manual pushing, pulling and twisting or torquing of meter-long concentric catheters and guidewires with pre-shaped tips (*1–3*). To advance inside microscopic brain vessels with sharp turns and loops, guidewires must be pushed against the vessel wall to slide forward, inherently carrying a risk of endothelial injury, perforation, dissection or vasospasm—the narrowing of blood vessels caused by the sudden contraction of the muscular vessel walls (*4–6*). The benchmark instruments on the market aim to mitigate these risks by exploiting distributed bending stiffness, shapable tips, and hydrophilic coatings (*7–9*). In parallel, steerable guidewires and catheters have been developed that allow active bending of the tip of the instrument inside the vessel using motorized tendons, hydraulics or magnetism (*10–20*). Recent work has shown that adding a helical protrusion to the outer surface of the instrument transforms rotation into advancement, further improving dexterity in endovascular interventions (*21*). Nevertheless, access remains constrained to arteries larger than 0.5 mm in diameter.

We have recently shown that microfabricated endovascular probes with a form factor of a ribbon can be transported by physiological fluid flow inside microfluidic vessel phantoms (*22*). Here we built upon this discovery and report a series of innovations in microrobotics that enable catheterization inside distal arteries as small as 180 μm in diameter and intraarterial infusion of medical liquids. Our in vivo results elevate flow-driven navigation concept into a viable clinical solution that can ultimately unlock new treatment avenues for neurological conditions, particularly to embolize distal cerebrovascular arteries of patients suffering from hemorrhage and hematoma expansion. Specifically, hypertensive patients are at risk of fatal intracerebral hemorrhagic stroke caused by occlusions or ruptures that appear in perforating arteries that are commonly sized at 50-400 μm, thus beyond the reach of existing microcatheter technology (*23–26*). Middle meningeal artery embolization has emerged as a minimally invasive strategy to treat chronic subdural hematoma (*27–28*). There are a number of ongoing randomized and multicenter clinical trials that aim to show the safety and efficacy of this endovascular treatment to eliminate the blood supply to the hematoma (*29–31*). Superselective embolization, which is challenging with the state-of-the-art microcatheters, is expected to improve the clinical outcomes by reducing the occurrence of reflux and off-target embolization (*32*).

# RESULTS

**Flow-driven magnetic microcatheter**

The flow-driven magnetic microcatheter (MagFlow) consists of four distinct parts: a magnetic radiopaque head that can be steered using external magnetic fields and visualized using X-ray fluoroscopy imaging, a microfabricated inflatable tube that can be transported by the arterial blood flow, a 1.8-meter-long reinforced tube that can be pushed by the operator, and a luer-lock hub at the proximal side for the infusion of fluids using a syringe (Fig. 1a). Different parts are assembled using fittings and a biocompatible glue, resulting in a water-tight ultraminiaturized catheter (Methods). To effectively harness hydrodynamic forces, the flexural rigidity of the microfabricated tube must be on the order of 1-10 pN m$^2$. To this end, thickness is the key geometric parameter as the bending stiffness of a rectangular beam scales cubically with its thickness. On the other hand, according to Poiseuille's Law, the flow rate is proportional to the fourth power of the lumen size. Therefore, decreasing the lumen size would drastically increase its flow resistance. We resolved this fundamental bottleneck in scaling by introducing a novel microcatheter design (Fig. 1b). The tube is kept in the original flat configuration until the target location is reached to harness the full potential of flow-driven navigation. Upon the initiation of infusion at the target location, the increase in the internal pressure inflates the tube, allowing for the injection of viscous fluids with minimal flow resistance. The key innovative step in the microfabrication process is the spatially selective bonding of 1.5 μm thick polyimide layers in a sandwich configuration, and patterning of an inner passivation layer that delaminates during the lumen formation (Methods and Supplementary Note 1). We designed MagFlow to be the first microcatheter deployable from a commercially available neurovascular microcatheter, providing superior distal arterial access (Fig. S1). MagFlow can be quickly deployed under saline perfusion (Methods and Fig. S2), a key component to keep the microfabricated tube under constant forward tension and avoid it to buckle or collapse when pushed by the reinforced tube. Upon entering the artery, viscous stresses applied by the physiological blood flow start pulling MagFlow downstream and keep it under tension throughout the operation, at which point saline perfusion can be terminated.

We microfabricated tubes in two different tip sizes, w$_{tip}$, from which full-length microcatheters are assembled for flow characterization and, ultimately, tested them during in vivo experiments (Fig. 1c). In all prototypes, the head diameter is the same as the tip size. Figures 1d,e show the cross-section of an inflated tube visualized by microcomputed tomography (Methods). At the inflated state, the tube section acquires a quasi-round configuration with a lumen area that can be calculated as $A =$ $w^2/\pi$ where $w$ is the width of the microfabricated tube at its flat state (Fig. S3). This equation can be used to define the width of the microfabricated tube considering the desired flow resistance during infusion. The flat tube inflates to the fullest extent when the input pressure is as little as 10 kPa. To assess clinical potential, we quantified the volumetric outflow rate with respect to input pressure (Figures 1f-h) for three different medically-relevant fluids of different viscosities, water (1 mPa s at 24 °C), non-adhesive liquid embolic agent (Squid from Balt, 12 mPa s at 37 °C) and contrast media (14.5 mPa s at 24 °C). As a reference, the suggested flow rate for microcatheter embolization using viscous glues is 0.16 mL min$^{-1}$ with a maximum of 0.3 mL min$^{-1}$ (*33*). We concluded that MagFlow is compatible with the infusion of both watery and viscous medical liquids including chemotherapy drugs, contrast media, and liquid embolic agents at clinically-relevant rates. We specifically chose Squid liquid embolic to prove the dimethyl sulfoxide (DMSO) compatibility of our microcatheter. Together with the other DMSO-based liquid embolic Onyx (Medtronic), they are widely used for neuroradiological interventions.

**Omnidirectional magnetic steering system**

Magnetic steering of endovascular instruments can be performed using permanent magnet(s) moving around the patient (*18*, *34-36*) or multiple electromagnets acting in concert (*37-40*). Inspired by a mechatronic device that can rotate a spherical permanent magnet about any axis of rotation using three omniwheels (*41*), we innovated on a hybrid magnetic steering system that is extremely compact and light-weight. The system consists of a spherical omnidirectional magnetic field generator (OmniMag) that is mounted on a collaborative robotic arm (Fig. 2a). In the center of OmniMag resides a cylindrical permanent magnet that can be oriented in any direction within a second (Fig. S4) using three orthogonal coils wounded around the magnet stator (Fig. 2b and Fig. S5). The peak magnetic field generated by the coils inside the stator (3.7 mT) at maximum current (2 A) applies a substantial torque (1.7 N m) on the permanent magnet as it has a strong magnetic dipole moment, $M$. The rotation of the permanent magnet changes the distribution of the magnetic field, **B**, around OmniMag, providing sufficient torque, $\tau$, to direct the head of the MagFlow with the magnetic dipole moment **m** towards the entry of the targeted daughter artery (Fig. 2b and Supplementary Note 2). Once the permanent magnet is rotated, it is maintained at this orientation until the system is instructed to change the magnetic field direction by the application of an idle current of 0.5 A to minimize energy consumption and heating. OmniMag does not move during catheter steering, and unlike clinical magnetic steering systems that consume kW of electric power and rely on active cooling systems to manage the heat generated by the coils, OmniMag runs on extremely low power (32 W during continuous actuation).

To validate the effectiveness of this steering strategy, we tested MagFlow in a continuously perfused plastic vessel model with bifurcations residing on a plane to facilitate visualization (Fig. 2c). We regulated the average fluid velocity according to the size of the vessels to remain in the physiologically relevant regime. We fixed the position of the OmniMag and steered MagFlow into different branches using a haptic stylus (Fig. 2d and Movie S1). From a clinical perspective, the fluoroscope c-arm, especially in the bi-plane configuration, poses stringent conditions regarding the positioning of magnets and their movement around the patient. We repeated the in vitro steering experiment for four alternative OmniMag positions around the phantom to show the versatility of our approach (Fig. S6). Table S1 reports the magnetic torque and force acting on MagFlow for all OmniMag configurations, which are on the order of nN m and nN, respectively. The OmniMag is programmed to maintain the orientation of the MagFlow head along a prescribed direction while being repositioned around the phantom (Movie S2)

**Navigating tortuous microvasculature in vivo**

To demonstrate the validity and efficacy of our technology in a clinical setting, we performed in vivo studies in a porcine model using real-time x-ray fluoroscope imaging. Porcine vessels in the head and spine region have comparable size, anatomy, and tortuosity to human cerebral vessels, thus providing an ideal testing platform (*42*, *43*). The first demonstration showcases MagFlow's unique ability to achieve distal access by navigating through narrow and tortuous arteries surrounding the spinal cord (Fig. 3a). In the beginning of each trial, a 5 Fr guide catheter was inserted through the femoral artery and advanced beyond the aortic arch into the vertebral artery (VA). Injection of iodine-based contrast agent from the guide-catheter enables the acquisition of a roadmap image mask of the distal arteries. A 2.7 Fr delivery microcatheter with an inner lumen of 0.7 mm was then deployed from the guide-catheter and navigated further inside the VA using a guidewire.

MagFlow was fed through the 2.7 Fr delivery microcatheter under continuous saline perfusion (5-25 mL min$^{-1}$) until its head was engaged with the bloodstream of the VA (Methods). First, we challenged MagFlow ($w_{tip}$ = 150 µm) to navigate inside the extremely tortuous radiculomedullary arteries that branch out from the VA, connecting to the anterior spinal artery (Fig. 3b). Measurements of contrast media perfusion over time during angiography showed that the flow velocity is between 0.8 and 2.8 cm s$^{-1}$ (Fig. S8 and Table S2). Notably, MagFlow could seamlessly navigate across an extreme 500 µm-diameter hair-pin turn with a curvature radius as small as 0.69 mm (Fig. 3c,d and Movie S3). As predicted by force measurements performed in biomimetic phantoms (*22*), we did not observe any sign of vasospasm in the angiograms during the in vivo experiments. Experienced neuroradiologists acknowledge that this would be extremely challenging, even with the smallest available microcatheter, the Magic 1.2 Fr (OD 0.4 mm) from Balt.

In the next trial, MagFlow ($w_{tip}$ = 250 µm) was deployed distally in the VA where several daughter muscular arteries branch off (Fig. 3e). After spatially referencing OmniMag with the fluoroscope frame (Methods), OmniMag was moved to a safe position outside the field of view of the fluoroscope using the controller of the robot arm (Fig. 3f and Fig. S9). The steering paradigm works as follows (see Supplementary Note 3 for details): looking at the displayed fluoroscope images of the radiopaque head and the vessels, the interventionalist handles an omnidirectional stylus that sets the desired $B$ direction around MagFlow. Knowing the catheter position with respect to the OmniMag (i.e. the vector $r$), the software asynchronously computes the orientation of $M$ encoding for a desired $B$, and energizes the coils to orient the magnet accordingly. We could repeatedly steer the microcatheter into six branches in less than a minute, where the last target artery sharply branches at orthogonal 90 ° angle from the VA (Fig. 3g and Movie S4). The average magnetic field strength during steering was 5.7 $\pm$ 1.4 mT, which corresponds to an estimated maximum magnetic torque of 16.9 $\pm$ 4.7 nN m (Fig. S10 and Table S3). We performed an additional navigation test within the infraorbital arteries (IOA) by repositioning the guide catheter in the external carotid artery (ECA) and the delivery microcatheter in the maxillary artery (MA) (Fig. S11). This time we intentionally positioned OmniMag inside the field of view of the fluoroscope to observe the rotation of the permanent magnet during in vivo navigation (Fig. S12 and Movie S5).

**Superselective embolization of the ophthalmic artery**

The major function of the microcatheter is the infusion of medical liquids inside the target arteries. We performed a final test to demonstrate that MagFlow is suited for superselective embolization—a technique that involves a local, slow, plug-and-push injection of a precipitating viscous embolic agent to occlude distal blood vessels and shunts such as in dural or pial arteriovenous malformations (*44*, *45*). Embolization leads to chronic ischemia, and thus tissue necrosis. It is therefore of paramount importance that clinicians can inject the drug locally, or superselectively, in order to avoid off-target effects. In pigs, the ophthalmic artery (OA) is a third-order bifurcation branch of the MA before it continues as the IOA (Fig. 4a). The terminal branches of the OA establish extensive anastomoses with branches of the facial, MA and superficial temporal arteries, providing ample opportunities to test distal navigation. Notably, the delivery microcatheter could not be inserted into the OA without creating vascular spasms at the ostium. MagFlow ($w_{tip}$ = 150 μm) was deployed in the ECA, and magnetically navigated into the OA and advanced more distally to one of the long ciliary arteries (CA), where the vessel diameter was comparable to the size of the 150μm-diameter MagFlow head (Fig. 4b, c and Movie S6).

Next, we demonstrated the feasibility of performing local angiography by manually injecting contrast medium directly into the OA using a 1 ml luer-lock syringe. Fig. 4d,e shows the deployment of MagFlow ($w_{tip}$ = 250 μm) from an anteroposterior view at a location following the first loop at the base of the OA. Contrast agent was injected at a rate of ~ 0.30 mL/min and successfully opacified the arteries downstream to the position of the MagFlow tip (Fig. 4f and Movie S7). In the following trial, we deployed MagFlow ($w_{tip}$ = 250 μm) in the OA with its tip as close as possible to the ocular globe within a long CA (Fig. 4g). A commercial embolization agent (Squid 12LD, Balt) was subsequently injected at a rate of 0.20 mL min$^{-1}$ under blank roadmap guidance until reflux was observed in order to form a proximal plug (Movie S8) (Methods). Once the proximal plug has formed, the embolic agent started to migrate downstream (Fig. 4h), completely blocking blood perfusion (Fig. 4i and Movie S9). Eventually, once the distal branches were embolized, reflux was visualized again, and the catheter was then removed before the embolic agent would migrate too proximally into non target territory to avoid collateral occlusion of normal branches.

**DISCUSSION**

Our results open a new avenue in medical procedures where ultraminiaturized instruments can be safely navigated far into very distal spinal and cerebral arteries currently unattainable using the conventional push-pull-torque paradigm. The flow-driven navigation technique would seamlessly work for even smaller MagFlow, which can be created following our microfabrication protocol. However, in vivo localization would become challenging due to the resolution limits of conventional angiographic systems. Our work may inspire future advancements in fluoroscopic imaging to enable the tracking of sub-100 µm radiopaque heads.

Superselective drug injection and embolization using MagFlow may satisfy current unmet needs, such as the superselective treatment of spinal and intracerebral hemorrhagic or oncological pathologies, a direction we will pursue in our future work. Alternatively, other organ systems could benefit from our technology, for example superselective prostate or vesical artery embolization in patients with invasive malignancies (*46-48*). In addition, our technology is ideally suited to navigate the miniaturized cardiovascular systems of children, pets and domestic animals, allowing pediatric, perhaps even foetal human and veterinary interventions. We demonstrated navigation and local delivery of a liquid contrast agent inside the OA. Intra-arterial chemotherapy is an established and effective treatment for retinoblastoma in children where a drug, melphalan, is injected from a microcatheter that is placed just at the ostium of the OA (*49*, *50*). Pushing the catheter tip past the ostium into the OA is expected to increase the efficacy of the treatment, however it is believed that this maneuver can cause ischemia in the majority of treated patients (*51*). Our technology could potentially allow locoregional injection of the drug directly into the OA without the complications associated with vasospasms or blocking of blood perfusion considering the significantly smaller diameter of MagFlow and gentleness of flow-driven navigation.

The principles of our flow-driven microcatheter may inspire the development of clinical instruments with alternative functions such as electronic probes for flow and pressure measurements (*22, 52*), and electrical neural recordings, and even perhaps far-reaching deep-brain stimulation or brain-computer implant technologies. Recent work has shown the feasibility of recording brain signals within arteries using ultraflexible microfabricated probes (*53,54*). This type of probes, currently deployed in a rudimentary manner, is readily applicable to our methodology for precise positioning inside microscopic human brain vasculature.

**MATERIALS AND METHODS**

**Microfabrication of the inflatable tube.** The microfabrication process aims to selectively bond interfaces between two layers of polyimide (PI) and create a virtual cavity within the ultra-thin flexible device using a sacrificial silicon dioxide ($SiO_2$) layer in between the bonded interfaces. Starting with a silicon wafer, we deposit and pattern layers of PI and $SiO_2$ using conventional techniques including spin coating, sputtering, photolithography, and dry etching, ensuring precise geometries and high-quality interfaces. Supplementary Note 1 provides the details of the microfabrication process. While the $SiO_2$ layer serves as an effective etch stop and sacrificial material, the PI layers offer excellent chemical resistance, facilitating the removal of the $SiO_2$. This methodology ensures compatible batch-processing of different materials using distinct chemistries. The 8 cm-long and 3 µm-thin tapered inflatable tubes are manually peeled from the wafer for the assembly of the microcatheter. The PI tube is compatible with liquid embolic agents such as Squid (Balt), PHIL (MicroVention Terumo) and Onyx (Medtronic) that uses Dimethyl sulfoxide (DMSO) as a solvent.

**Manufacturing of the magnetic head.** The radiopaque magnetic head (Fig. S13) was fabricated by casting a platinum/Iridium (Pt/Ir) ring (L0.5 mm long and OD/ID Ø150/90 µm or OD/ID Ø250/190 µm, iMC Intertech) inside a composite of 5 µm Neodymium Iron Boron (NdFeB) magnetic particles (Magnequench) and polydimethylsiloxane (4:1 wt%; Sylgard 184, Dow-Corning). The magnetic composite was cured at 65°C for at least 2h. The polymerized head was then magnetized at a field strength of 3500 $kAm^{-1}$ using an impulse magnetizer (Magnet-Physik).

**Assembly of full-length MagFlow.** MagFlow is assembled from medical-grade components compatible with DMSO solvent. A proximal luer hub made of polyethylene (Vygon) was bonded to a custom polyimide tube reinforced with a metal coil (Microlumen) with a total length of 1.8 m, OD/ID Ø470/ Ø368 µm. A smaller polyimide tube (Microlumen) was inserted and bonded to create a tapered connection (2 cm-long, OD/ID Ø330/Ø203 µm), which facilitates assembly with the inflatable microfabricated tube. The microfabricated tube was designed in such a way that it could be attached to this smaller polyimide tube at its proximal side and the magnetic head on its distal side (Fig. S14). To bond each element a primer (SF 770, Loctite) was used prior to applying the instant glue (406, Loctite) with a pressure dispenser (Ultimus, Nordson). The assembled rigid tube was then tightly fitted into the proximal end of the inflatable microtube. A radiopaque Pt/Ir marker ring (OD/ID Ø450/Ø350 µm, L0.3 mm) (Prince&Izant) was then slid over and glued with a medical grade UV curable glue (3301, Loctite). The magnetic head was attached to the tip of the inflatable catheter using UV-curable glue (3301, Loctite) and micropositioners.

**Microcomputer tomography.** Microcomputer tomography (CT) scans of the flow-driven microcatheter were done using a machine (UltraTom, RX Solutions) that performs 1800 scans per revolution at setpoint parameters of 60 kV and 50 µA. A 400 nm focal spot size allows reaching a voxel resolution of 495 nm. Sample preparation was performed by filling the microcatheter with a UV-curable glue (NOA81, Thorlabs), which is rapidly cured upon UV exposure to visualize the tube in its inflated state.

**Flow measurements.** Flow characterizations were performed by perfusing the flow-driven microcatheters at isobaric conditions using a pressure dispenser (Ultimus, Nordson). We measured the total mass ejected from the flow-driven microcatheter over a fixed time interval, between 30 s and 2 min according to the diameter of the tested device (n = 3). The volumetric flow rate is the total mass divided by the mass density of the liquid. We repeated the tests with three different liquids: water, iodine contrast medium (Iomeron 350 mg mL$^{-1}$, Bracco) and Ethylene Vinyl Alcohol Copolymer (EVOH) embolic liquid dissolved in DMSO (Squid 12 LD, Balt). Flow rates for water and contrast agent were performed at room temperature. The contrast agent was used as received without further dilution. Tests with the embolic liquid (Squid 12 LD, Balt) were performed in a glovebox (Terra Universal) at 37 °C. The embolic liquid was prepared by vortexing the vial for 20 min at 2,500 rpm at room temperature, which was then heated to 37 °C. The microcatheter was flushed with pure DMSO before the infusion of the embolic liquid.

**Magnetic steering system.** The magnetic navigation system is composed of a robotic arm (UR16e, Universal Robots) and a custom-built omnidirectional magnetic end effector. The end effector includes a 3D ball bearing system with an inner rotor hosting a high grade (magnetic remanence Br = 1.47 T) NdFeB cylindrical (L80 mm x Ø80 mm) permanent magnet (Vacodym 722 HR, Vacuumschmelze) that can be oriented in any direction using a set of 3 orthogonal electromagnetic coils (enamelled copper wire Ø0.56 mm, Multicomp Pro) wound around the bearing cage (stator) in a spherical configuration. The current inside the coils is regulated with a digital multiaxis amplifier (MiniMACS, Maxon). The maximum current per coil is set at 2 A, corresponding to a current density of 8.1 A mm$^{-2}$. The current density allows long-term dynamic control of magnet orientation without active cooling. The holder, the rotor and the stator of the omnidirectional end effector are machined from non-magnetic aluminium material. The frame within the stator and rotor that holds the polyoxymethylene (POM) balls (Ø3 mm, Maagtechnic) is 3D printed (J55, Stratasys) from polymethyl methacrylate (PMMA). The reconstruction of the direction of the magnet was performed using the Moore-Penrose pseudoinverse as described elsewhere[33]. Magnetic field strength is monitored by two on-board hall sensors (TLE493D, Infineon) connected to a digital I/O acquisition card (Arduino Mega). The user interface is a haptic stylus (Touch, 3D Systems) that defines the desired magnetic field direction at the microcatheter head. This information is used to compute the corresponding orientation of the permanent magnet. The control software, written in Python programming language (Version 3.10.11), allows asynchronous operations and communications with different interfaces at a global computing cycle rate of 10 Hz.

**In vitro experiments.** Vessel phantoms drawn with CAD software (Solidworks 2022) were printed from ABS (Protolabs). The fluid flow was provided by a peristaltic pump (LP- BT100-2J, Drifton). The oscillations in pressure were dampened using a Windkessel element. A solution of 42.5 wt% of glycerol (Sigma-Aldrich) in deionized water was used as a viscosity-matching blood analogue. The flow-driven microcatheter is introduced into the channels through a 2.7 Fr microcatheter (see in vivo methods). Visualization was performed with a Nikon D7200 camera.

**In vivo catheterization protocol.** In vivo experiments were performed at the Veranex France animal facility in Paris by an experienced interventional neuroradiologist with more than 15 years of neurointerventional practice and in vivo studies (P.J.M). All experiments were approved by the EPFL Animal Research Ethics Committee (AREC000018) and the French Ministry of Research (APAFIS 12534-2017121118586862). Five female pigs (31 ± 3 kg) were anesthetized by the veterinary team, and vital parameters including electrical activity of the heart, blood pressure, end tidal CO2, blood oxygen level and core body temperature were continuously monitored and recorded throughout the acute procedures. At the end of each trial, the pig was euthanized. The femoral arteries were cannulated with a 5 Fr introducer sheath (Radiofocus, Terumo) using the Seldinger technique to deploy a 5 Fr guide catheter (Envoy, Codman) inside the external carotid artery or vertebral artery under X-ray fluoroscopy guidance (ALLURA Xper FD 20F, Philips Healthcare). Then, a 2.7 Fr delivery microcatheter (Excelsior XT27, Stryker or Marksman, Medtronic or Rebar-27, Medtronic) was navigated using a 0.014" guidewire (Traxcess 14, Microvention) to the distal portion of the vertebral artery or branching arteries of the external carotid artery.

**In vivo calibration of the magnetic navigation system**. See Supplementary Note 3 for the details of the calibration protocol. Briefly, the robot was positioned at a location next to the operation table while leaving ample room for the rotation of the C-arm fluoroscope. The position of the robot with respect to the fluoroscope coordinates was recorded in the navigation system software. To extract the catheter position, an intraoperative angiographic CT-scan of the target arteries was performed. The 3D position of the catheter in the fluoroscope coordinates was intraoperatively identified by measuring its position using 3D orthogonal multiplanar reconstruction.

**In vivo deployment and navigation of MagFlow**. MagFlow was deployed from the 2.7 Fr microcatheter by manual feeding at 1 to 5 cm s$^{-1}$ under continuous perfusion of a saline solution at a flow rate of 5 to 25 mL min$^{-1}$ to ensure that the inflatable microtube would not buckle. Perfusion was provided by a 20 mL syringe through a 3-way valve (Fig. S15). Once the radiopaque head of MagFlow enter the artery, the forces applied by the blood flow were sufficient to assist manual feeding of the catheter, at which point the saline perfusion was stopped. The navigation and magnetic steering of MagFlow was performed under fluoroscopic imaging using roadmaps in different pig vessels, which were generated using digital subtraction angiography technique. During the in vivo trials, the system was operated in an entirely open-loop fashion. The surgeon looked at the screen and instructed the system to bend the MagFlow head in a certain direction using the haptic stylus. The system used this information and calculated the orientation of OmniMag that would generate magnetic fields capable of bending the MagFlow head along the chosen direction.

**In vivo superselective embolization and angiography**. Embolic liquid that consists of DMSO-dissolved ethylene–vinyl alcohol (EVOH) co-polymer (Squid 12 LD, Balt) was prepared by mixing the vial at 2,500 rpm for 20 min using a vortex machine (Thermomixer, Eppendorf). Once MagFlow was positioned within the target artery, it was first flushed with 3 mL of physiological saline solution, followed by a rinse with 0.8 mL pure DMSO. The embolic liquid was pre-filled in a 1 mL syringe and immediately used after mixing to avoid the sedimentation of the micronized tantalum particles to maintain homogeneous radio-opacity during the subsequent embolization and follow its distal migration as well as its proximal reflux. The solvent DMSO dissipates into the blood, causing the copolymer EVOH to precipitate in situ into a spongy, coherent embolus.

**Image processing.** X-Ray fluoroscopy and CT images were analyzed and reconstructed volumetrically with Osirix MD (v14.0.1) and subsequently processed with ImageJ2 (v2.9.0). Projected angiographies from the fluoroscope were resized from 512x512 to 4096x4096 pixels using a bilinear interpolation. Micro CT images were resized and pre-processed with a moving median average filter (5 pixels radius). CT volumetric reconstructions were performed with Volume Viewer plugin.

**Acknowledgements**

This work was supported by the European Research Council (ERC) under the European Union's Horizon 2020 research and innovation program (899792) and Innosuisse (104.465.1 IP-ENG). L.P. acknowledge funding from BRIDGE Proof of Concept (40B1-0_209182) and Innogrant (S133-IN-23-06). M.A.G. acknowledge funding from Swiss National Science Foundation MD-PhD program (323530_214536). We thank Dr. Nicolas Borenstein and the Veranex France team for their exceptional support during the animal work preparation and experimentation. We thank PIXE facility for their assistance with the microCT imaging and Balt for providing the SQUID 12 LD embolic liquid. We thank Justina Venckute for her contributions to the microfabrication of the inflatable tube, Julian Raub for his assistance with the in vivo tests, and Pierre Oppliger for his work on the programming of OmniMag.

**Author Contributions**

L.P., P.J.M. and M.S.S designed the in vivo experiments, L.P., E.O., M.A.G., and P.J.M. performed the in vivo experiments and analyzed the data, L.P. and M.A.G. developed the manufacturing process for the microcatheters, L.P. and E.O. constructed and programmed the magnetic steering system, L.P. and M.S.S. wrote the manuscript with contributions from all authors, M.S.S. supervised the research. All authors contributed to the discussion of the results and the manuscript revision.

**Competing Interest**

L.P. and M.S.S. filed a patent on the ultraflexible flow directed device and system (EP3999161A1 and US20220265969A1) and another patent on the magnetic guide system (EP23199238.9). The authors declare no other competing interests.

**Data and materials availability**

All data are available in the main text or the supplementary materials.

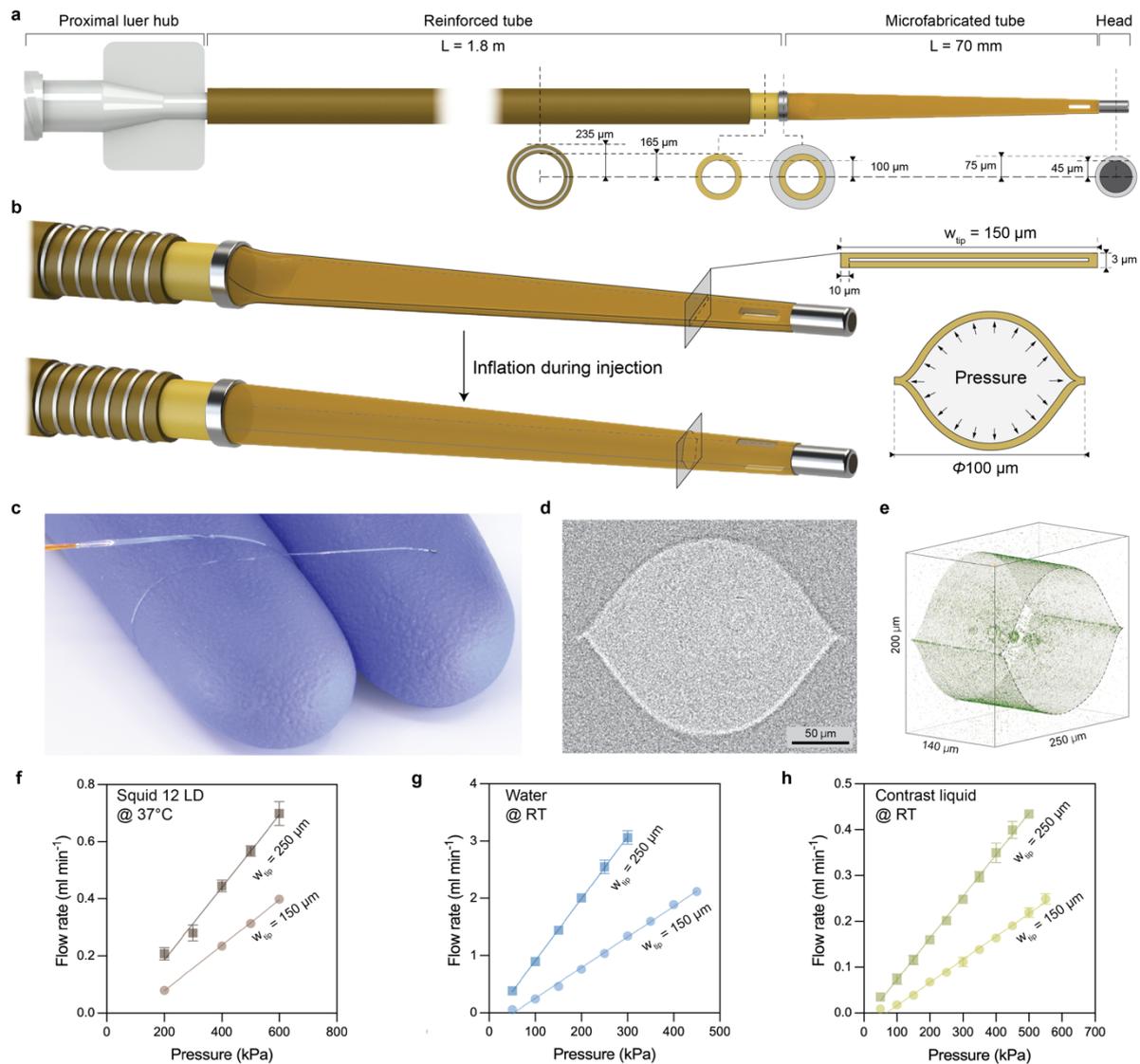

**Fig. 1. MagFlow: flow-driven magnetic microcatheter**. **a** MagFlow microcatheter is an assembly of four parts: the proximal polyethylene luer-lock hub to connect the microcatheter to a syringe for the injection of medical fluids, 1.8 m-long polyimide tube reinforced with steel braids, 70 mm-long ultraflexible microfabricated inflatable tube with two outlets at the distal tip, and a radiopaque magnetic head. The cross-sectional dimensions are shown below the MagFlow. **b** A magnified view of MagFlow highlighting the geometric transformation of the microfabricated tube during injection of medical fluids. The two outlets at the distal end allow for intra-arterial liquid ejection. The cross-sectional view shows the dimensions of the lumen at the distal tip before and after inflation. The distal width, $w_{tip}$, measures 150 µm at rest, corresponding to the head diameter. The width reduces to 100 µm during inflation while the tube expands in the orthogonal axis. **c** Optical image of MagFlow wrapped around human fingers. **d** Microcomputed tomography image of an inflated microfabricated tube along with (**e**) volumetric reconstruction of several images. **f-h** The flow rate at the outlet of MagFlow versus input fluid pressure for two different $w_{tip}$. From left to right, liquid embolic agent Squid 12 LD (Balt) (at 37 °C), water at room temperature, and contrast liquid (Iomeron 350 mg ml$^{-1}$, Bracco) at room temperature, respectively. The test for the embolic agent was done at body temperature considering the change in viscosity of the compound with temperature. Data are generated from the average of 3 devices (n = 3) and error bars indicate one standard deviation.

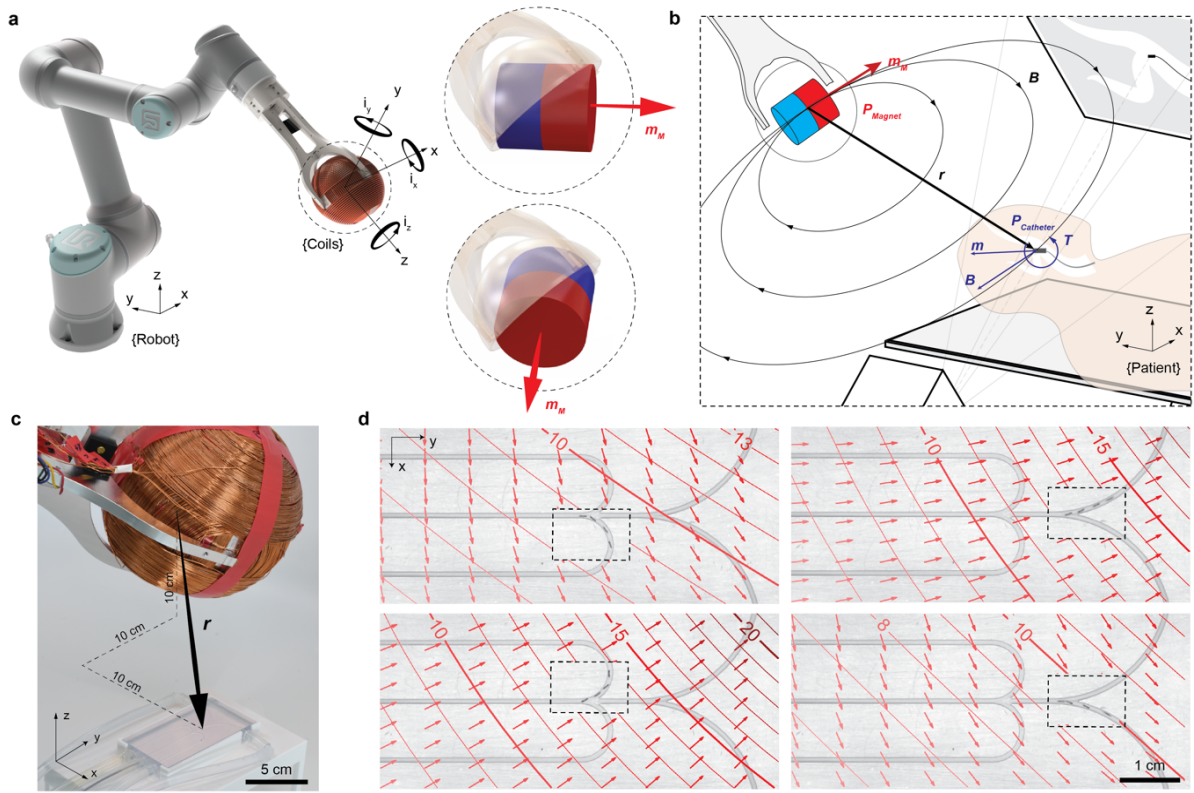

**Fig. 2. OmniMag: omnidirectional magnetic steering system**. **a** The robotic steering platform consists of a collaborative robotic arm and the end-effector, OmniMag. Energization of the 3 orthogonal coils reorient a large permanent magnet, with magnetic dipole $M$, fixed within an omnidirectional rotor. **b** Magnetic field $B$, emanating from the permanent magnet applies a torque $\tau$ on MagFlow magnetic head, with magnetic dipole $m$, from a distance $|r|$. **c** Positioning of MagFlow during in vitro navigation in a biomimetic fluidic phantom. The position vector, $r$, that describes the position of the MagFlow head with respect to the centre of the permanent magnet is chosen as $r = [10 \quad -10 \quad -10]$ cm. **d** Top view of the fluidic phantom showing superposed time-lapsed images of MagFlow head inside branching vessels. Red arrows and lines indicate the direction of the magnetic field and the field magnitude (in mT), respectively. See Fig. S6 and Table S1 for detailed measurements.

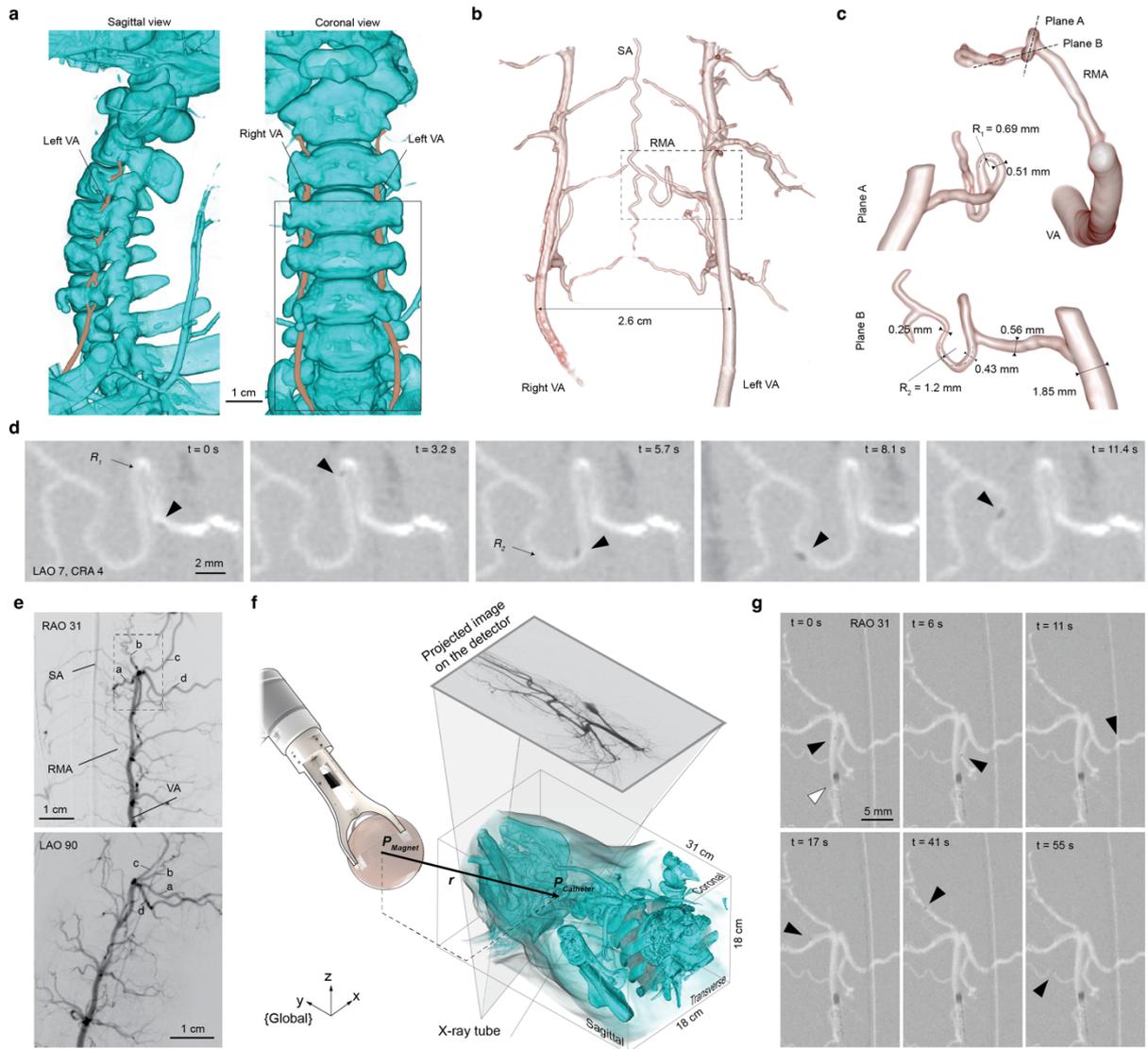

**Fig. 3. Flow-driven navigation and magnetic steering inside tortuous distal arteries. a** Sagittal and coronal view of the volumetric reconstruction of the VAs (red) passing through the vertebras (blue). See Fig. S7 for the description of the body planes. Boxed in solid line corresponds to the region where MagFlow was navigated in the spine vasculature. **b** Reconstruction of the main spinal vessels, the spinal artery (SA) and the vertebral artery (VA), that are highlighted with the rectangular box in (**a**). Dashed box shows the radiculomedullary artery (RMA) catheterized by MagFlow. **c** Isolated reconstruction of the RMA originating from the left VA illustrating the extreme tortuosity and miniature size of the two consecutive hair-pins with curvature radii of $R_1$ = 0.69 mm and $R_2$ = 1.2 mm and diameter decreasing from 0.56 mm proximally to 0.25 mm at the SA anastomoses. **d** Time-lapse fluoroscopy images of MagFlow navigating across the RMA hair-pins. **e** Quasi-AP (top) and lateral (bottom) view angiography of the distal VA branching off into muscular arteries where MagFlow was magnetically steered. The target arteries are denoted by letters a-d according to the chronological order. **f** Not-in-scale schematic illustration of the positioning of OmniMag with respect to the animal along with the fluoroscope field-of-view. The position vector is $r = [14.6 \quad -14.9 \quad -4.4]$ cm, corresponding to a magnetic field strength $|B|$ at the MagFlow head between 4.9 mT (radial field) and 9.7 mT (axial field). **g** Time-lapse fluoroscope

images of MagFlow head inside the muscular arteries branching from the distal VA. All black arrows indicate MagFlow head position and white arrow indicates tip position of the 2.7 Fr delivery microcatheter.

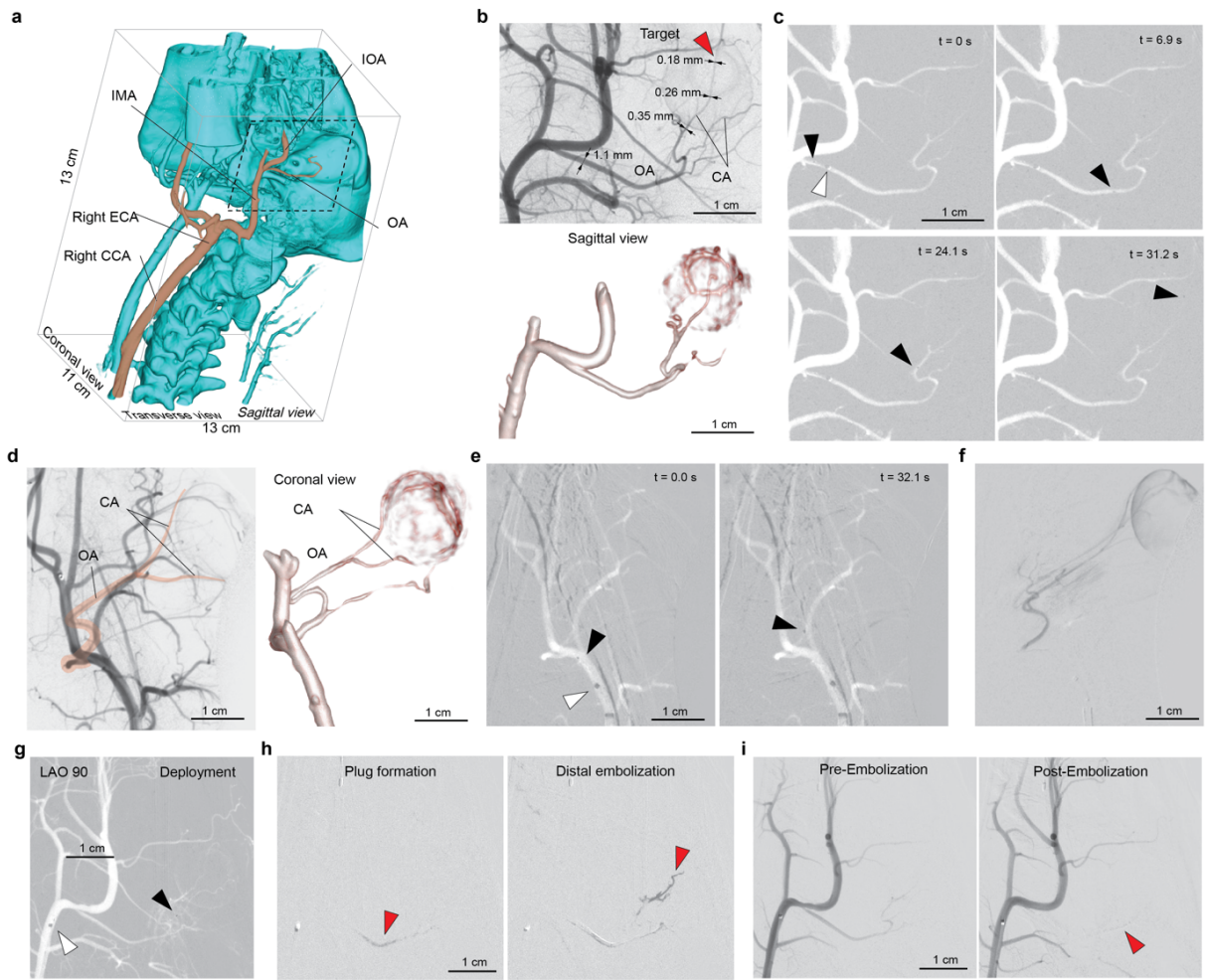

**Fig. 4. In vivo superselective arterial embolization. a** Digital CT reconstruction of the neck-head vasculature (red) highlighting the ophthalmic artery (OA) and its parent arteries. **b** Sagittal view of the roadmap (top) and volumetric reconstruction (bottom) of the OA and the ciliary artery (CA) surrounding the ocular globe. Red arrow indicates the target artery measuring Ø0.18 mm in diameter. **c** Time-lapse fluoroscope images of MagFlow navigation from the OA ostium to the distal ciliary artery. **d** Coronal view of the roadmap (left, with fake colours) and volumetric reconstruction of the OA. **e** Snapshots of the distal deployment of MagFlow in the OA. **f** Fluoroscopic image during focal injection of contrast agent from MagFlow. **g** Sagittal roadmap of the OA and positioning of MagFlow in the distal OA. **h** Superselective embolization of the OA and CA through embolic plug formation (left) allowing embolization of the distal arteries (right), shown with red arrows. **i** Comparative angiography images before and after embolization showing complete obstruction of the OA and distal branches (red arrow). All black arrows indicate MagFlow head position.